\documentclass[aps,prb,preprint,groupedaddress,nobibnotes]{revtex4-2}

\usepackage{amssymb}
\usepackage{epstopdf}
\usepackage{graphicx}
\usepackage[caption=false]{subfig}
\usepackage{float}
\usepackage{amsmath}
\usepackage{dcolumn}
\usepackage{color}
\usepackage{booktabs}
\usepackage[pdftex,colorlinks=true,linkcolor=red,citecolor=blue]{hyperref}
\usepackage{soul}

\begin{document}

\title{Magnetic properties and Fermi-surface topology of kagome magnet HoV$_6$Sn$_6$}
\author{Tucker Beekmann$^1$}
\author{Ganesh Pokharel$^{2}$}
\author{Wyatt Edward Ackerman$^{2}$}
\author{Stephen D. Wilson$^{3}$}
\author{David E. Graf$^{4,5}$}
\author{Keshav Shrestha$^1$}
\email{Corresponding e-mail: kshrestha@wtamu.edu}

\affiliation{$^{1}$Department of Chemistry and Physics, West Texas A$\&$M University, Canyon, Texas 79016, USA}
\affiliation{$^{2}$Perry College of Mathematics, Computing, and Sciences, University of West Georgia, Carrollton, Georgia 30118, USA}
\affiliation{$^{3}$Materials Department, University of California, Santa Barbara, California 93106, USA}
\affiliation{$^{4}$National High Magnetic Field Laboratory, Tallahassee, Florida 32310, USA}
\affiliation{$^{5}$Department of Physics, Florida State University, Tallahassee, Florida 32306, USA}

\begin{abstract}
We report a study of the magnetic properties and Fermi-surface topology of the kagome magnet HoV$_6$Sn$_6$ using magnetization and tunnel-diode-oscillator (TDO) measurements. The temperature-dependent magnetization exhibits Curie-Weiss behavior at high temperatures and reveals long-range magnetic order below $T_c \approx 2.4$ K. Below $T_c$, the field-dependent magnetization reveals pronounced magnetic anisotropy, with saturation moments of $\approx$ 9.5 and 7~$\mu_B$/Ho for fields applied along the out-of-plane and in-plane directions, respectively. To probe the Fermi surface, we perform TDO measurements in magnetic fields up to 41.5~T and at temperatures down to 0.35~K. The TDO signal exhibits well-defined quantum oscillations with multiple frequencies extending up to $\sim$9~kT. Angular- and temperature-dependent measurements are used to determine the Fermi-surface geometry and cyclotron effective masses, respectively. To elucidate the electronic structure, we perform first-principles calculations of the electronic bands, density of states (DOS), and Fermi surface. The calculated band structure exhibits flat-band features, van Hove singularities, and Dirac-like crossings near the Fermi level. The DOS near the Fermi level is dominated by the V-$d$ orbitals, indicating that the V kagome layers play a major role in determining the low-energy electronic properties of HoV$_6$Sn$_6$. Our calculations further support a ferromagnetic ground state with the Ho moments oriented along the $c$ axis, consistent with neutron-scattering measurements. The calculated quantum-oscillation frequencies from the Fermi-surface pockets of HoV$_6$Sn$_6$ are in overall good agreement with the experimentally observed frequencies. Our combined experimental and theoretical results provide insight into the magnetic anisotropy and electronic structure of HoV$_6$Sn$_6$ and reveal its multiband Fermi-surface topology.
\end{abstract}


\maketitle
\section{Introduction}
Kagome materials, characterized by a lattice of corner-sharing triangles, have attracted considerable interest because the interplay between lattice geometry, electronic structure, and interactions can give rise to a variety of unconventional quantum phenomena~\cite{wilson2024v3sb5,yin2022topological,wang2023quantum}. In particular, the kagome lattice can host geometrical frustration, nontrivial band topology, flat electronic bands, and van Hove singularities, providing a versatile platform for exploring correlated and topological states of matter~\cite{di2026kagome,hu2023electronic,wang2024topological,jiang2023kagome,bernevig2022progress}. These characteristics have been associated with a diverse range of emergent phenomena, including charge-density-wave (CDW) order, unconventional superconductivity, anomalous Hall effects, and complex magnetic states~\cite{werhahn2022kagome,ortiz2023evolution,ortiz2024intricate,ye2018massive,liu2018giant,li2025giant}.

One important class of kagome materials is the $AB_6X_6$ family, commonly referred to as the 166 family~\cite{pokharel2021electronic,shi2023topological,lee2022anisotropic,pervin2026interplay}. This family offers considerable chemical flexibility, with the $A$ site accommodating a variety of alkali, alkaline-earth, and rare-earth elements, including Li, Mg, Yb, Sm, and Gd, while the $B$ site is typically occupied by a transition metal such as Co, Cr, Mn, V, or Ni. The $X$ site is generally occupied by a group-IV element, such as Si, Ge, or Sn. This chemical tunability gives rise to a rich variety of electronic and magnetic ground states, particularly in compounds containing magnetic kagome lattices. For example, a spin-polarized Dirac cone has been reported in YMn$_6$Sn$_6$~\cite{li2021dirac}, large anomalous Hall effects have been observed in LiMn$_6$Sn$_6$ and GdMn$_6$Sn$_6$~\cite{chen2021large,asaba2020anomalous}, and Chern topological magnetism has been reported in TbMn$_6$Sn$_6$~\cite{yin2020quantum}. In addition, YMn$_6$Sn$_6$ exhibits competing magnetic phases, highlighting the sensitivity of the electronic and magnetic properties to chemical composition and magnetic interactions~\cite{ghimire2020competing}. GdV$_6$Sn$_6$, HoV$_6$Sn$_6$, and YV$_6$Sn$_6$ have also been reported to exhibit topological electronic features~\cite{peng2021realizing,hu2022tunable}.

In contrast to the $A$V$_3$Sb$_5$ ($A$ = K, Cs, Rb) family, in which CDW order and superconductivity are widely observed~\cite{ortiz2019new,ortiz2020cs,shrestha2023fermi,shrestha2022nontrivial}, CDW order appears to be less common within the 166 family. ScV$_6$Sn$_6$ was the first member of the 166 family in which a CDW transition was experimentally established, occurring at approximately 92 K~\cite{arachchige2022charge,zheng2024quantum,meier2023tiny}. More recently, LuNb$_6$Sn$_6$, which is isostructural to ScV$_6$Sn$_6$, was also found to exhibit CDW order below 85 K~\cite{ortiz2025stability,beekmann2026fermiology,lou2025orbital,ingham2024theory}. Interestingly, both ScV$_6$Sn$_6$ and LuNb$_6$Sn$_6$ undergo a $\sqrt{3}\times\sqrt{3}\times3$ structural modulation in their CDW phases~\cite{ortiz2025stability,beekmann2026electronic,arachchige2022charge}.Despite extensive studies of the 116 family, previous work has largely focused on its magnetic properties, while its fermiology remains comparatively unexplored.

Motivated by the rich electronic and magnetic properties of the 166 family, we investigate the kagome magnet HoV$_6$Sn$_6$ using magnetization, tunnel-diode oscillator (TDO) measurements, and first-principles calculations. Although electrical transport, magnetotransport, magnetic, and heat-capacity measurements have been reported~\cite{devi2026coexistence,zeng2024magnetic,lee2022anisotropic,zhang2022electronic} for HoV$_6$Sn$_6$ to understand its magnetic properties and ground state, to the best of our knowledge, quantum-oscillation studies of this compound have not yet been reported. HoV$_6$Sn$_6$ crystallizes in a structure containing V-based kagome layers, as shown in Fig.~\ref{Fig1}(a). We observe a magnetic transition at $T_c\approx2.4$ K and an anisotropic ferromagnetic-like ground state. High-field TDO measurements reveal quantum oscillations with frequencies extending up to $\sim9$ kT, providing access to the Fermi-surface topology and charge-carrier effective masses. The experimentally observed quantum-oscillation frequencies are in good agreement with those calculated from the extremal cross sections of the Fermi surface, establishing a direct correspondence between the measured quantum oscillations and the underlying electronic structure.

\section{Experimental and computational details}
Single crystals of HoV$_6$Sn$_6$ were grown via a conventional flux-based growth technique. Ho (pieces, 99.9\%), Sn (shot, 99.99\%), V (pieces, 99.7\%) were loaded inside a Canfield crucible set with a molar ratio of 1:6:20 and then sealed in a quartz tube filled with 1/3 Ar-atmosphere. The sealed tubes with a mixture of elemental ingredients were heated at 1125$^o$C at a heating rate of 200$^o$C/hr and then dwelled for 12 hours. Then, the mixture was slowly cooled down to 780 $^o$C at a rate of 2$^o$C/h to promote crystal growth. Below 760 $^o$C, binary V–Sn phases become more stable compared to the ternary phase. Plate-like single crystals were separated from the molten flux via centrifuging at 780 $^o$C before binary phase start appearing. The crystals grown by this technique were usually a few millimeters in length and $\sim$ 200 $\mu m$ in thickness. The surfaces of some of these crystals were contaminated with weak tin flux. So, they were subsequently cleaned with dilute HCl and Q-tips to remove the surface contamination.

Magnetization measurements were performed using a Quantum Design Magnetic Properties Measurement System (MPMS). Quantum-oscillation measurements were performed using a TDO circuit, in which the sample was placed inside the coil of a self-resonating LC circuit~\cite{van1975tunnel,coffey2000measuring}. The resonant frequency of the circuit is sensitive to changes in the magnetic susceptibility of the sample. Measurements were performed at temperatures down to 0.35 K in magnetic fields up to 41.5 T at the National High Magnetic Field Laboratory, Tallahassee. The field-dependent TDO signal was analyzed to extract the quantum-oscillation frequencies and their temperature dependence.

First-principles calculations were performed using both WIEN2k~\cite{blaha2020wien2k,blaha2001wien2k} and the open-source computational package Quantum ESPRESSO (QE)~\cite{giannozzi2009quantum,giannozzi2017advanced,giannozzi2020quantum}. The crystal structure was initially optimized using the \textit{vc-relax} scheme in QE with conjugate-gradient (CG) diagonalization. Subsequent electronic-structure calculations, including the electronic band structure and Fermi surface, were performed using WIEN2k. The exchange-correlation energy was treated within the generalized gradient approximation (GGA) using the Perdew-Burke-Ernzerhof (PBE) functional~\cite{perdew1996generalized}. A Hubbard $U$ of 6~eV was applied to the Ho-$4f$ states to account for the localized nature of the Ho-$4f$ electrons. To obtain high-resolution Fermi-surface maps, a dense $25\times25\times12$ Monkhorst--Pack $\mathbf{k}$-point mesh was employed. Theoretical quantum-oscillation frequencies were calculated from the extremal Fermi-surface cross sections using the \textsc{SKEAF} code~\cite{julian2012numerical}.

\section{Results and Discussion}
\subsection{Magnetic properties}
\begin{figure*}
\centering
\includegraphics[width=1.0\linewidth]{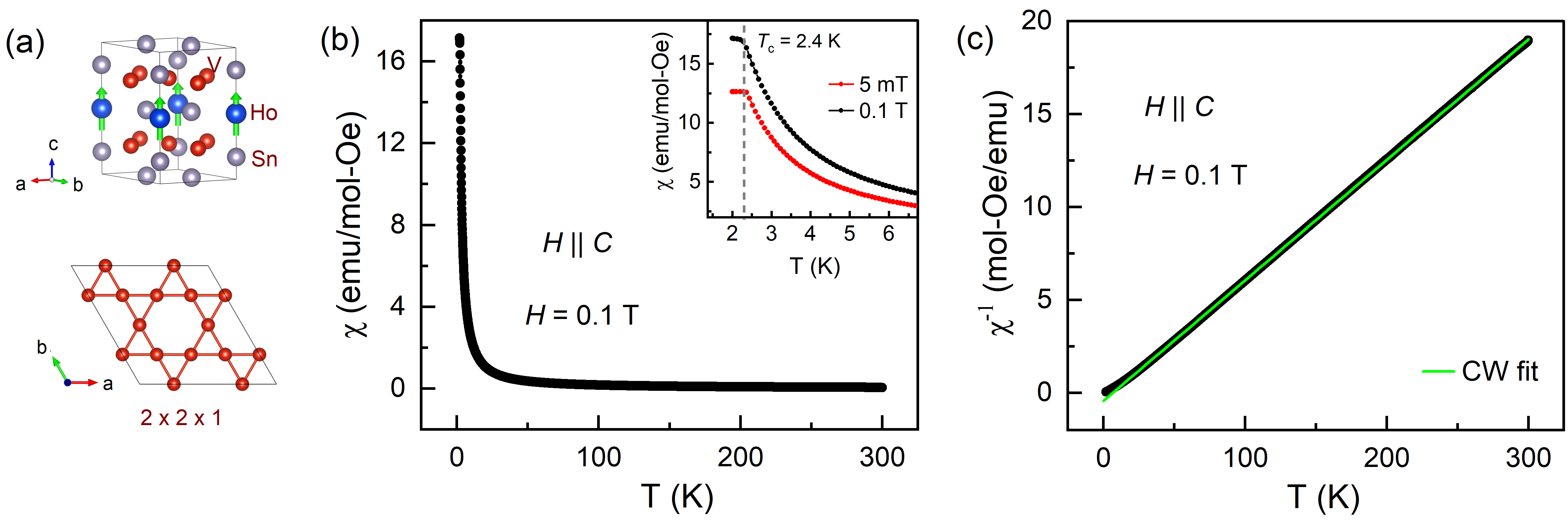}
\caption{(a) Unit cell of HoV$_6$Sn$_6$ (upper panel). The arrow indicates the magnetic moment direction of the Ho atoms. The bottom panel shows the top view of a $2\times2\times1$ supercell, highlighting the two-dimensional kagome network formed by the V atoms. (b) Temperature dependence of the magnetic susceptibility measured under an applied magnetic field of 0.1 T along the $c$ axis. The inset shows the low-temperature magnetic susceptibility measured under applied fields of 0.1 T and 5 mT, highlighting the magnetic transition. (c) Temperature dependence of the inverse magnetic susceptibility. The solid line represents the best fit to the Curie--Weiss law.}\label{Fig1}
\end{figure*}

Figure~\ref{Fig1}(b) shows the temperature dependence of the magnetic susceptibility, $\chi(T)$, of a HoV$_6$Sn$_6$ single crystal measured in an applied magnetic field of 0.1 T with $H\parallel c$. At high temperatures, $\chi(T)$ exhibits Curie-Weiss-like behavior, while a clear anomaly develops below $T_c\sim2.4$ K, indicating the onset of long-range magnetic order, as shown in the inset. Previous bulk measurements, including electrical transport, heat capacity, and magnetization, have reported magnetic transitions at 2.3 K~\cite{lee2022anisotropic}, 2.4 K~\cite{devi2026coexistence}, 2.7 K~\cite{zeng2024magnetic},  and 3.71 K~\cite{zhang2022electronic} in HoV$_6$Sn$_6$. These values are consistent with the magnetic transition temperature observed in our measurements.

To determine the magnetic parameters, the inverse susceptibility, $\chi^{-1}(T)$, is plotted as a function of temperature in Fig.~\ref{Fig1}(c). The high-temperature susceptibility is described by the Curie-Weiss law~\cite{kittel2018introduction},
$$
\chi(T)=\frac{C}{T-\theta_{\rm CW}},
$$
where $C$ is the Curie constant and $\theta_{\rm CW}$ is the Weiss temperature. The solid line in Fig.~\ref{Fig1}(c) represents a fit to the experimental data using the Curie-Weiss expression. The good agreement between the experimental data and the fit indicates that the high-temperature susceptibility is well described by the Curie-Weiss behavior. The fit yields a Curie constant of $C=15.427(1)$~emu~K~mol$^{-1}$~Oe$^{-1}$ and a Weiss temperature of $\theta_{\rm CW}=6.63(1)$~K. These fitted parameters are in good agreement with previously reported values~\cite{zeng2024magnetic,lee2022anisotropic,zhang2022electronic} for HoV$_6$Sn$_6$. Furthermore, the corresponding effective magnetic moment is $\mu_{\rm eff}=11.11(1)\,\mu_B$/f.u., which is close to the expected free-ion value of $10.61\,\mu_B$ for Ho$^{3+}$.

\begin{figure}
\centering
\includegraphics[width=1.0\linewidth]{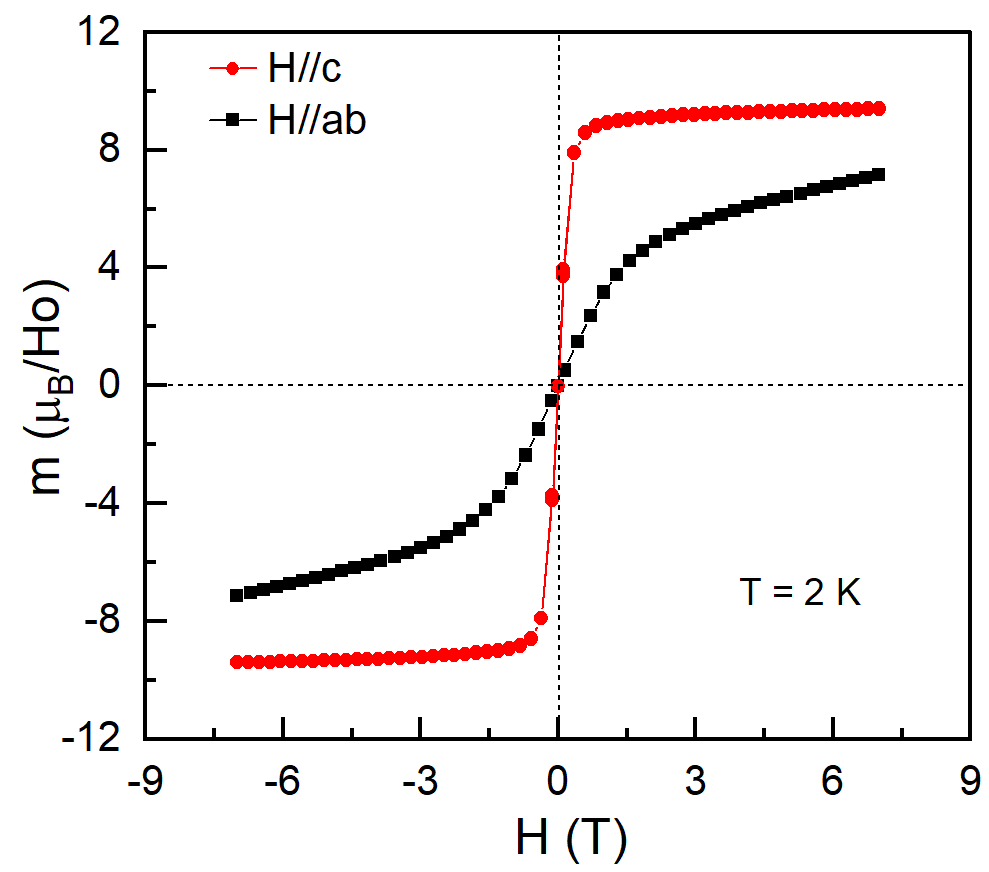}
\caption{Field-dependent magnetization $M(H)$ of HoV$_6$Sn$_6$ measured at $T=2$~K with the magnetic field applied along the out-of-plane direction ($H\parallel c$) and within the $ab$ plane ($H\parallel ab$). The magnetization exhibits pronounced anisotropy, with a rapid approach to a saturation moment of $9.5\,\mu_B$/Ho near 1~T for $H\parallel c$, whereas the magnetization for $H\parallel ab$ increases gradually without clear saturation up to 7~T.}\label{Fig2}
\end{figure}

To further characterize the magnetic transition near $T_c\sim2.4$ K, we measured the field-dependent magnetization, $M(H)$, at $T=2$ K with the magnetic field applied along both the $c$ axis and within the $ab$ plane. As shown in Fig.~\ref{Fig2}, the $M(H)$ curves exhibit pronounced magnetic anisotropy. For $H\parallel c$, the magnetization increases rapidly with increasing field and approaches saturation above $\sim1$ T, reaching a saturation moment of approximately $9.5\,\mu_B$/Ho. In contrast, for $H\parallel ab$, the magnetization increases more gradually and reaches approximately $7.2\,\mu_B$/Ho at 7 T, without showing clear saturation within the measured field range. No appreciable hysteresis is observed for either field orientation, indicating a small coercive field. The strong anisotropy, together with the larger saturation moment for $H\parallel c$, indicates that the Ho magnetic moments preferentially align along the $c$ axis. This observation is consistent with recent neutron-scattering measurements, which also indicate that the magnetic moments in HoV$_6$Sn$_6$ are oriented predominantly along the $c$ axis~\cite{zhou2024ground}. Similar anisotropic field-dependent magnetization and comparable saturation moments have been reported for HoV$_6$Sn$_6$ by other groups~\cite{zeng2024magnetic,lee2022anisotropic,zhang2022electronic} and have also been observed in other members of the $R$V$_6$Sn$_6$ family, including $R=$ Tb, Dy, and Er~\cite{zhou2024ground,pokharel2022highly}.

\subsection{Tunnel-diode oscillator measurements}
\begin{figure*}
\centering
\includegraphics[width=1.0\linewidth]{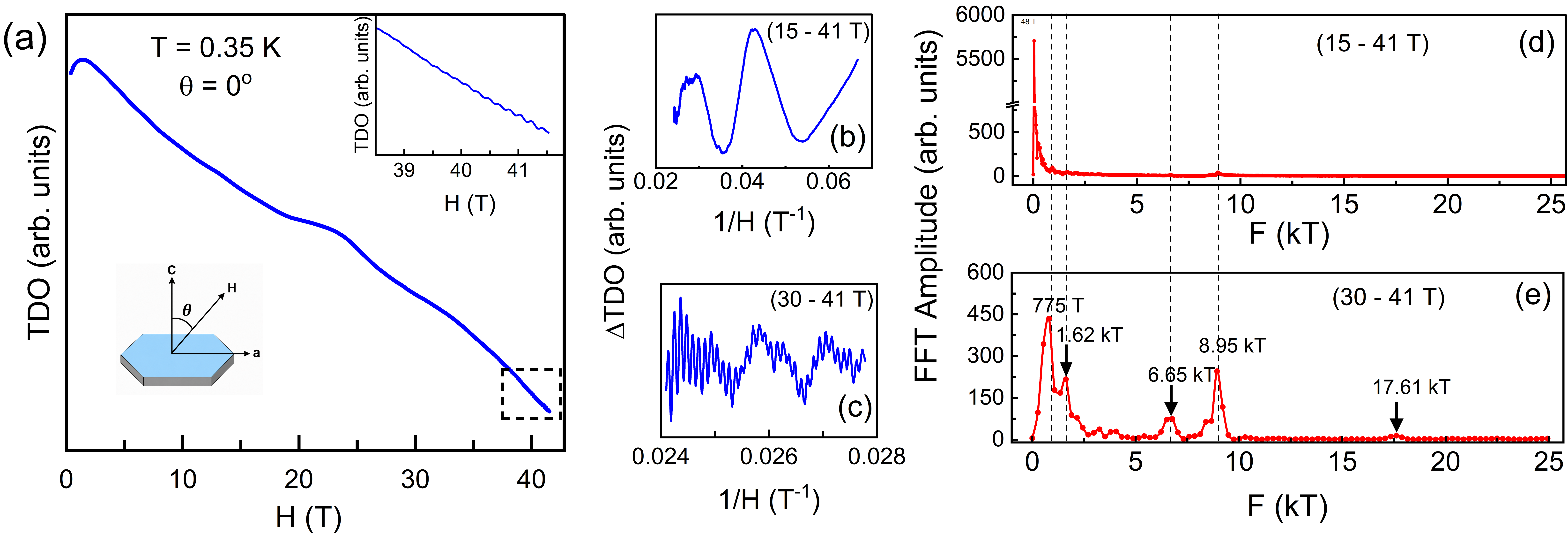}
\caption{(a) TDO signal as a function of magnetic field for HoV$_6$Sn$_6$ at $T=0.35$ K and $\theta=0^\circ$, measured up to 41.5 T. Low-frequency quantum oscillations become apparent above 10 T, while a high-frequency oscillation emerges above 30 T, as highlighted in the upper inset. Lower inset: Schematic diagram illustrating the tilt angle, $\theta$. Background-subtracted TDO signals over the magnetic-field ranges of (b) 15–41.5 T and (c) 30–41.5 T, revealing clear modulations associated with the low- and high-frequency quantum oscillations. (d) and (e) Fourier spectra obtained from the background-subtracted signals in (b) and (c), respectively. Six prominent frequencies are observed at 48 T, 775 T, 1.62 kT, 6.65 kT, 8.95 kT, and 17.61 kT. The vertical dashed lines are guides to the eye.}\label{Fig3}
\end{figure*}

To investigate the Fermi-surface properties of this material, we performed high-field measurements using the TDO technique in magnetic fields up to 41.5 T. This technique has been successfully employed to probe the Fermi surfaces of a variety of quantum and kagome materials~\cite{rehfuss2024quantum,shrestha2022nontrivial,broyles2022effect,wu2025fermi}. Figure~\ref{Fig3}(a) shows the field dependence of the TDO signal measured at $T=0.35$ K for $\theta=0^\circ$, where $\theta$ is defined as the angle between the sample surface and the $c$ axis, as illustrated in the inset of Fig.~\ref{Fig3}(a). The TDO signal exhibits clear signatures of quantum oscillations above 10 T. In addition to the dominant low-frequency oscillations, a high-frequency signal is also present at fields above $30$ T within the region indicated by the dashed box in Fig.~\ref{Fig3}(a), where the oscillations are more clearly resolved in the enlarged view shown in the inset.

To further analyze the oscillatory components, we subtracted the slowly varying background from the TDO signal. The background was obtained by fitting the field-dependent TDO signal to a third-order polynomial and subsequently subtracting the fitted contribution. Because the amplitude of the high-frequency oscillations is substantially smaller than that of the low-frequency component, we analyzed two different magnetic-field windows, $15$--$41.5$ T and $30$--$41.5$ T, as shown in Figs.~\ref{Fig3}(b) and \ref{Fig3}(c), respectively. In the broader field window [Fig.~\ref{Fig3}(b)], the oscillatory signal is dominated by the low-frequency component, whereas restricting the analysis to the higher-field window [Fig.~\ref{Fig3}(c)] significantly enhances the visibility of the high-frequency oscillations.

We then performed fast Fourier-transform (FFT) analysis of the background-subtracted signals to determine the corresponding oscillation frequencies. The resulting FFT spectra are presented in Figs.~\ref{Fig3}(d) and \ref{Fig3}(e) for the $15$--$41.5$ T and $30$--$41.5$ T field windows, respectively. The FFT spectrum obtained from the broader field window is dominated by a low frequency of $48$ T. In contrast, the high-field FFT spectrum reveals several additional frequencies at $775$ T, $1.62$ kT, $6.65$ kT, $8.95$ kT, and $17.61$ kT. These high-frequency components are also present in the FFT spectrum obtained from the $15$--$41.5$ T window, but their amplitudes are strongly suppressed relative to the dominant $48$ T peak. The corresponding frequencies are indicated by the vertical dashed lines in Fig.~\ref{Fig3}(d). The presence of high-frequency signals reaching up to 9 kT has also been reported in other 166 kagome compounds~\cite{dhital2024fermi,shtefiienko2025electronic,beekmann2026electronic,phillips2025electrical,shrestha2023electronic} and in the $A$V$_3$Sb$_5$ kagome family~\cite{ortiz2021fermi,zhang2022emergence,broyles2022effect}.

To further investigate the Fermi-surface topology, we performed angle-dependent TDO measurements by rotating the magnetic field relative to the sample plane. The quantum-oscillation frequency is related to the extremal cross-sectional area of the Fermi surface through the Onsager relation~\cite{kittel2018introduction,miertschin2024anisotropic}, $F=\frac{\hbar}{2\pi e}A_F,$ where $F$ is the quantum-oscillation frequency, $A_F$ is the extremal cross-sectional area of the Fermi surface perpendicular to the applied magnetic field, $\hbar$ is the reduced Planck constant, and $e$ is the elementary charge. Thus, tracking the angular dependence of the oscillation frequencies provides a direct probe of the Fermi-surface geometry~\cite{ando2013topological}. To map this angular dependence, we rotated the magnetic field from the $c$-axis toward the $a$-axis within the $ac$ plane.

\begin{figure}
\centering
\includegraphics[width=1.0\linewidth]{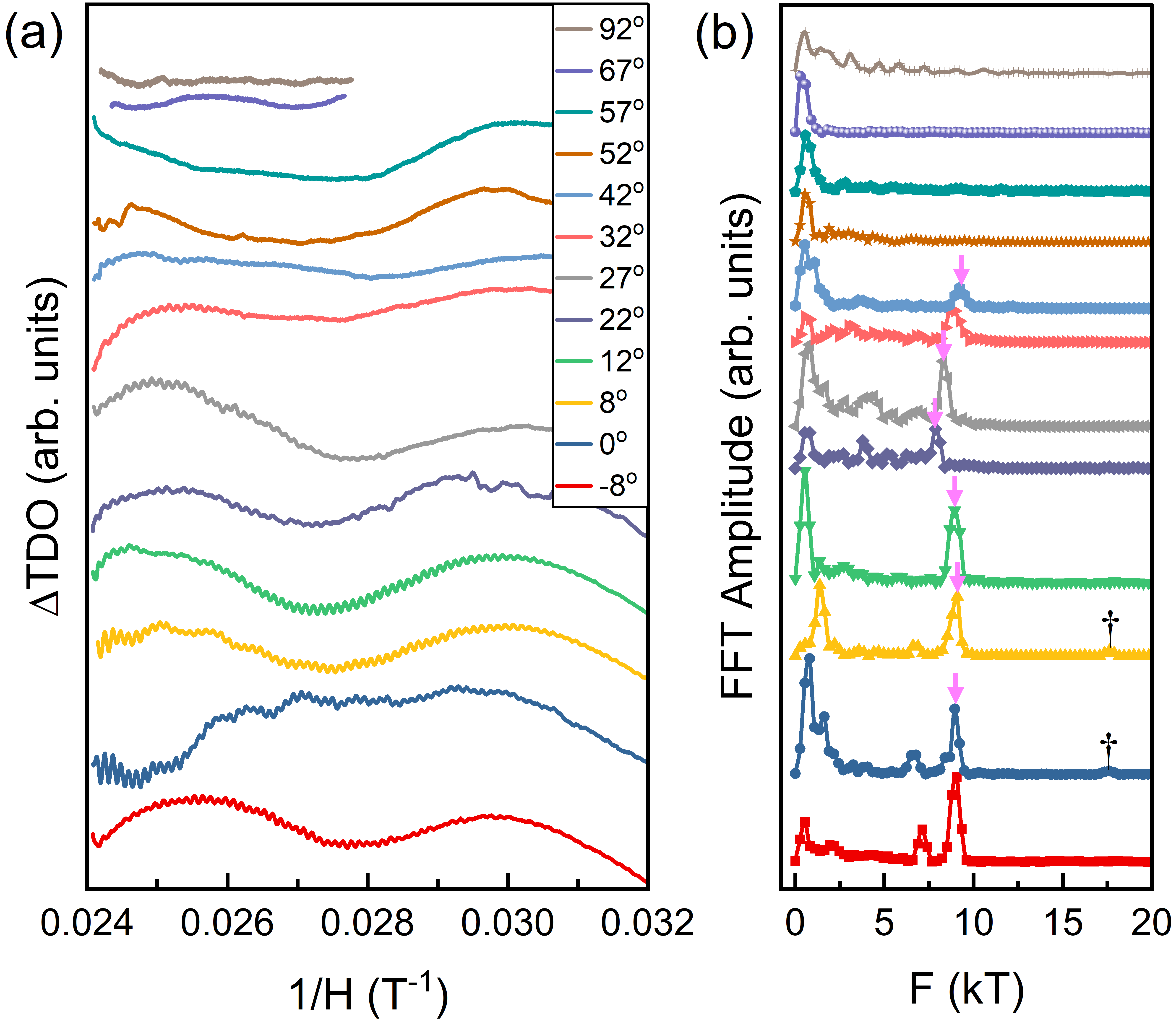}
\caption{(a) Background-subtracted TDO signals at selected tilt angles, $\theta$. At low $\theta$, a low-frequency oscillation is clearly resolved, with the high-frequency oscillation superimposed on it. The high-frequency component becomes progressively weaker and is no longer resolved at larger $\theta$. (b) Corresponding frequency spectra obtained from the oscillations in (a). The high-frequency component near 8.9 kT, indicated by the arrow, is resolved up to $\theta = 42^\circ$ but disappears at larger $\theta$. The low-frequency component near 800 T persists over the entire measured angular range. A higher-frequency component near 18 kT, marked by the dagger, is resolved only at a few low-$\theta$ angles. The curves in (a) and (b) are shifted vertically for better visibility.}\label{Fig4}  
\end{figure}

Figure~\ref{Fig4}(a) shows the background-subtracted TDO signal as a function of magnetic field for representative tilt angles $\theta$. Here, $\theta$ is the angle between the $c$ axis and the magnetic-field direction, as illustrated in the lower inset of Fig.~\ref{Fig3}(a). Both low- and high-frequency oscillatory components can be resolved over a range of angles, with their oscillation periods evolving systematically as the field orientation is varied. The high-frequency oscillations become progressively weaker with increasing $\theta$ and are no longer clearly resolved above approximately $42^\circ$. The evolution of the oscillation frequencies with field orientation is more clearly revealed by the corresponding FFT spectra shown in Fig.~\ref{Fig4}(b).

The low-frequency component persists over the entire angular range investigated, up to $\theta=92^\circ$, with its FFT peak remaining nearly unchanged in frequency. In contrast, the high-frequency components exhibit a pronounced angular dependence. In particular, the feature near $8.95$ kT, indicated by the arrow in Fig.~\ref{Fig4}(b), can be resolved up to $\theta\approx42^\circ$ and shifts systematically toward higher frequencies with increasing $\theta$. The distinct angular evolution of the low- and high-frequency components indicates substantially different Fermi-surface geometries associated with the corresponding extremal orbits. The second high-frequency feature at $\sim17.61$ kT, marked by the dagger symbol, is resolved at only two measured angles. Its frequency is nearly twice that of the $\sim8.95$ kT feature, indicated by the arrow, suggesting that it is the second harmonic of the $8.95$ kT feature.

\begin{figure}
\centering
\includegraphics[width=1.0\linewidth]{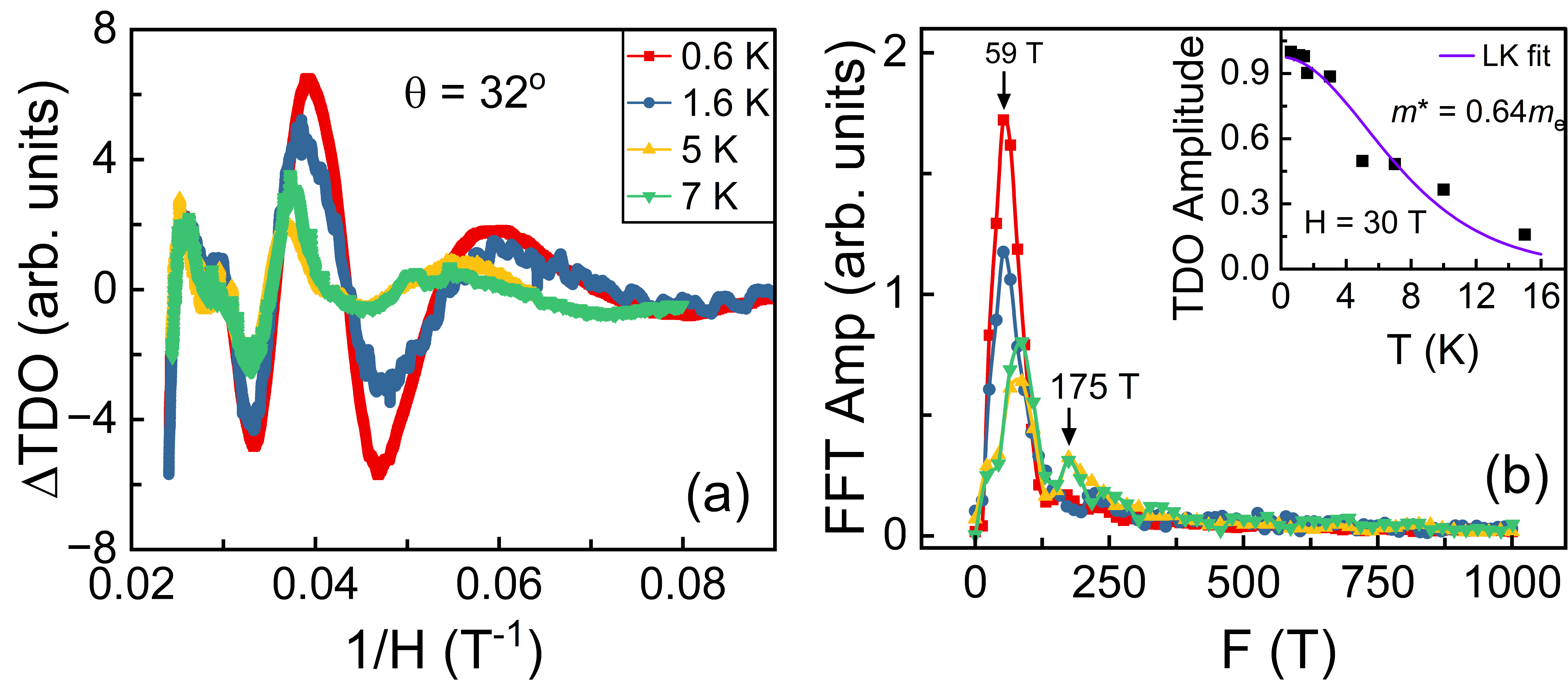}
\caption{(a) Temperature dependence of the TDO signal of HoV$_6$Sn$_6$ at selected temperatures, measured at $\theta = 32^\circ$. The oscillation amplitude decreases with increasing temperature. (b) FFT spectra of the quantum oscillations shown in (a). A prominent peak is observed at 59 T, along with a small peak at 175 T. Inset: Temperature dependence of the TDO oscillation amplitude at $H = 30$ T. The solid line represents the best fit to the LK formula~\ref{LK}.}\label{Fig5}
\end{figure}

To determine the effective mass of the charge carriers, we performed TDO measurements at several temperatures with the magnetic field oriented at $\theta=32^\circ$. The quantum oscillations were extracted by subtracting a smooth polynomial background from the TDO signal. The resulting oscillations at selected temperatures are shown in Fig.~\ref{Fig5}(a). At this field orientation, only a low-frequency oscillation with $F_\alpha=59$~T is observed, as evident from the FFT spectra in Fig.~\ref{Fig5}(b). A small peak is observed at approximately twice the frequency of $F_\alpha$, corresponding to the second harmonic of $F_\alpha$. The oscillation amplitude decreases with increasing temperature, consistent with the thermal damping described by the Lifshitz--Kosevich (LK) formalism~\cite{shoenberg2009magnetic}. According to the LK theory, the temperature- and field-dependent oscillation amplitude is given by

\begin{equation}
\Delta \mathrm{TDO}(T,H) \propto
e^{-\lambda_D(H)}
\frac{\lambda(T/H)}
{\sinh[\lambda(T/H)]},
\label{LK}
\end{equation}

where

$$\lambda_D(H)=
\frac{2\pi^2 k_B}{\hbar e}
m^*\frac{T_D}{H},$$

and

$$\lambda(T/H)=
\frac{2\pi^2 k_B}{\hbar e}
m^*\frac{T}{H}.$$

Here, $T_D$, $k_B$, and $m^*$ are the Dingle temperature, Boltzmann constant, and quasiparticle effective mass, respectively. The first term in Eq.~\ref{LK} is the Dingle factor, which describes the damping associated with impurity scattering, while the second term describes the thermal damping of the oscillation amplitude due to the thermal broadening of the Fermi surface. To extract the effective mass, we analyzed the temperature dependence of the oscillation amplitude at $H=30$~T, as shown in the inset of Fig.~\ref{Fig5}(b). The measured amplitude is well described by the thermal damping term of the LK expression. The fit yields an effective mass of $m^* = 0.64(6)m_e$, where $m_e$ is the free-electron mass. This value falls within the range of effective masses reported for low-frequency quantum oscillations in other 166 kagome compounds, including $m^* = 0.36(1)m_e$ at 95 T in GdV$_6$Sn$_6$~\cite{dhital2024fermi}, $0.19m_e$ at 77 T in ScV$_6$Sn$_6$~\cite{shrestha2023electronic}, $0.19m_e$ at 12 T in LuV$_6$Sn$_6$~\cite{phillips2025electrical}, and $(0.10$–$0.14)m_e$ for frequencies between 35 and 96 T in RMn$_6$Sn$_6$ ($R=$ Gd–Tm, Lu)~\cite{ma2021rare}.

\subsection{Electronic band structure and Fermi surface}
To gain further insight into the experimentally observed quantum oscillations and the underlying electronic structure, we performed first-principles calculations for HoV$_6$Sn$_6$. The compound develops long-range magnetic order below $\sim 2.4$ K, and our $M$-$H$ measurements [Fig.~\ref{Fig2}] indicate a ferromagnetic ground state, consistent with previous neutron-scattering results showing ferromagnetic ordering with the Ho moments aligned along the crystallographic $c$ axis~\cite{zhou2024ground}. We performed DFT+\(U\) calculations for both nonmagnetic and ferromagnetic configurations using \(U=6\) eV for the Ho \(4f\) states, consistent with previous DFT+\(U\) studies of Ho-based compounds~\cite{lopez2025effect,guo20253d}. The ferromagnetic configuration is lower in energy than the nonmagnetic configuration by $\sim 4.5$ eV. We therefore calculated the electronic bands and Fermi surface for the ferromagnetic configuration with the Ho moments oriented along the $c$ axis.

\begin{figure*}
\centering
\includegraphics[width=1.0\linewidth]{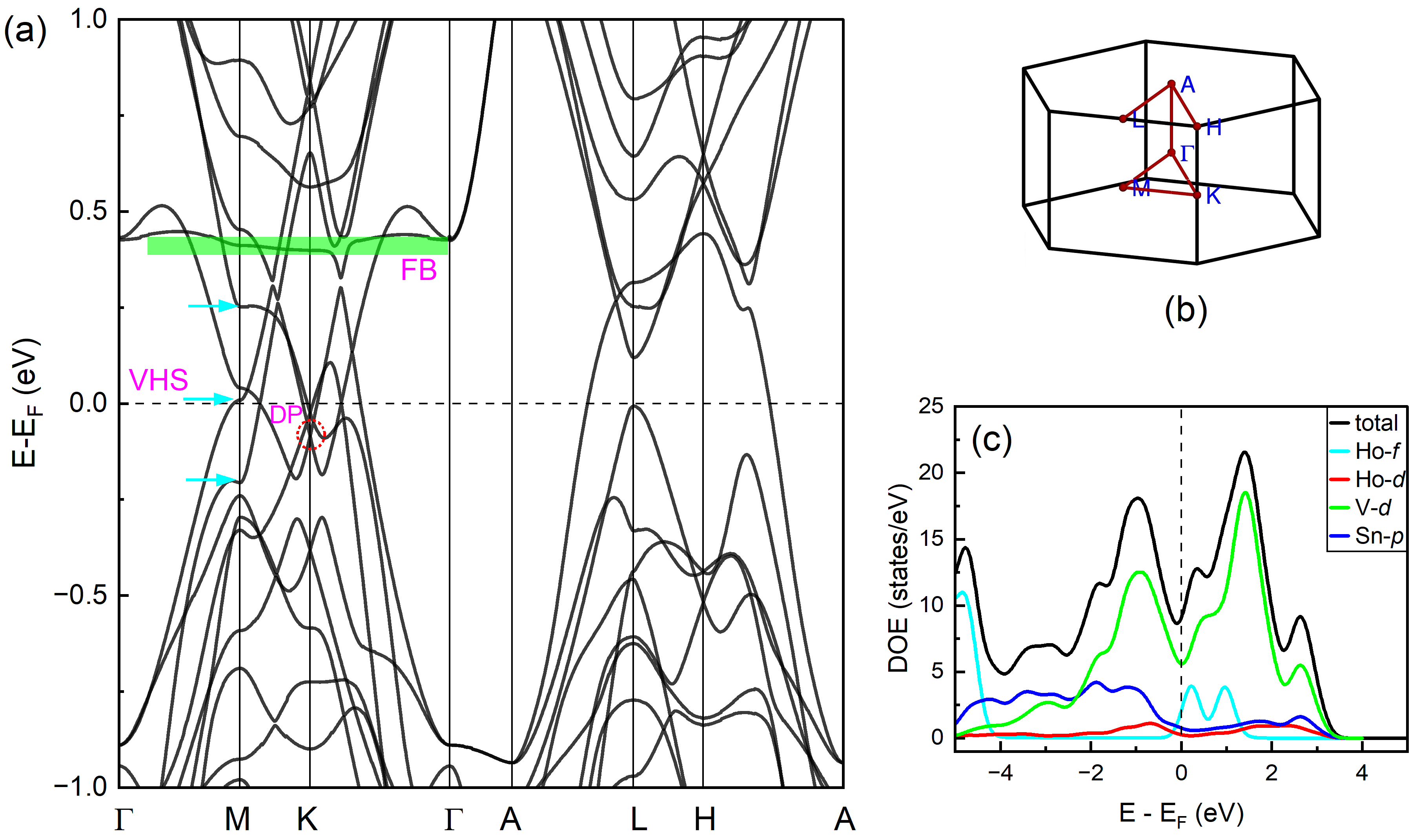}
\caption{(a) Electronic band structure of HoV$_6$Sn$_6$ calculated without spin-orbit coupling (SOC) along the high-symmetry $k$-path. Several notable features, including a flat band, van Hove singularities, and Dirac points, are present near the Fermi level, as indicated by the shaded region, arrow, and dotted circle, respectively. (b) Brillouin zone of HoV$_6$Sn$_6$ showing the high-symmetry $k$-path used for the band-structure calculation. (c) Atom- and orbital-resolved density of states (DOS) of HoV$_6$Sn$_6$. The DOS near the Fermi level is predominantly derived from the V-$d$ orbitals.}\label{Fig6}
\end{figure*}

Figure~\ref{Fig6}(a) shows the calculated electronic band structure of HoV$_6$Sn$_6$ without spin-orbit coupling (SOC) along the selected high-symmetry $k$ path shown in Fig.~\ref{Fig6}(b). The electronic band features of HoV$_6$Sn$_6$ are similar to those of other kagome compounds~\cite{hu2024magnetic, sakhya2024diverse, bhandari2024first, bhandari2025pressure,rosenberg2024probing}, with the presence of several characteristic features associated with kagome-derived electronic structures, including a relatively flat band (FB), van Hove singularities (VHSs), and Dirac points (DPs), highlighted by the shaded region, arrow, and dotted circle, respectively. Notably, several of these features occur relatively close to the Fermi level. Multiple VHSs are present near the $M$ point, with one VHS located very close to the Fermi level. The relatively flat band is located $\sim$0.4 eV above the Fermi level, while the Dirac point at the $K$ point lies $\sim$60--70 meV below the Fermi level. In addition, several Dirac-like band crossings are observed along other high-symmetry directions; for example, a crossing along the $K$-$\Gamma$ direction occurs close to the Fermi level. The coexistence of these flat bands, VHSs, and Dirac-like crossings over an energy range near the Fermi level highlights the rich electronic structure of HoV$_6$Sn$_6$ and provides an important basis for understanding its magnetic and quantum-oscillation properties.

Figure~\ref{Fig6}(c) presents the atom- and orbital-resolved density of states (DOS) of HoV$_6$Sn$_6$ in the vicinity of the Fermi level. The total DOS near the Fermi level is dominated by the V-$d$ states, while the Ho-$f$ orbitals provide a smaller but finite contribution. In contrast, the contributions from the Ho-$p$ and Sn-$p$ orbitals are comparatively weak in this energy range. Since the V atoms form the kagome layers in HoV$_6$Sn$_6$, the dominant V-$d$ character of the DOS indicates that the V kagome network plays a central role in determining the low-energy electronic structure of the compound. Furthermore, the finite total DOS at the Fermi level is consistent with the metallic character of HoV$_6$Sn$_6$. This behavior is also evident from the calculated band structure in Fig.~\ref{Fig6}(a), where multiple electronic bands cross the Fermi level. The agreement between the finite DOS at the Fermi level and the band crossings therefore provides a consistent picture of the metallic electronic state of HoV$_6$Sn$_6$, with the low-energy states predominantly derived from the V-$d$ orbitals.

\begin{figure}
\centering
\includegraphics[width=1.0\linewidth]{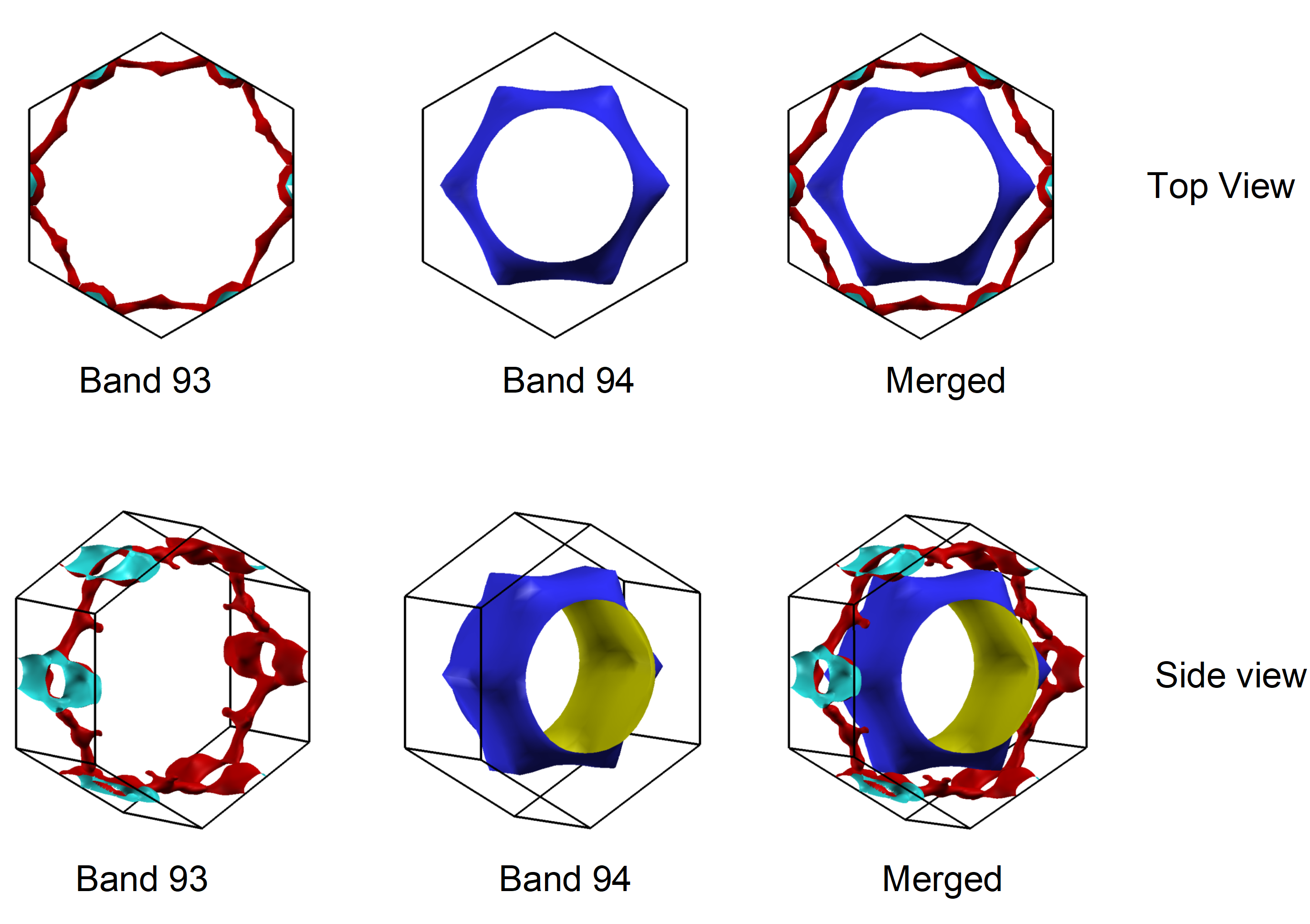}
\caption{Top and side views of the band-resolved Fermi surface of HoV$_6$Sn$_6$. Bands 93 and 94 cross the Fermi level and contribute to the Fermi surface. Band 93 exhibits a ring-like feature along the boundary of the Brillouin zone, whereas Band 94 forms a barrel-shaped Fermi surface centered at the $\Gamma$ point. The individual panels show the Fermi-surface contributions from Bands 93 and 94, while the last subfigure in each of the upper and lower panels shows the combined Fermi surface from both bands.}\label{Fig7}
\end{figure}

Two bands, Band 93 and Band 94, cross the Fermi level in HoV$_6$Sn$_6$. Their corresponding Fermi surfaces are shown in Fig.~\ref{Fig7}. Band 93 forms a complex, strongly corrugated, and interconnected Fermi-surface sheet with multiple necks and pocket-like regions, reflecting a pronounced three-dimensional topology. Band 94, on the other hand, forms a predominantly barrel-shaped sheet centered at the $\Gamma$ point. The combined Fermi surface, shown in the rightmost panel, consists of these two distinct sheets. Such Fermi-surface features, including interconnected Fermi sheets at the Brillouin-zone boundary and a barrel-shaped Fermi pocket, have also been observed in LuV$_6$Sn$_6$~\cite{phillips2025electrical}, GdV$_6$Sn$_6$~\cite{dhital2024fermi}, and YV$_6$Sn$_6$~\cite{shtefiienko2025electronic,rosenberg2024probing}. 

\subsection{Comparison of experimental and calculated frequencies}
\begin{figure}
\centering
\includegraphics[width=1.0\linewidth]{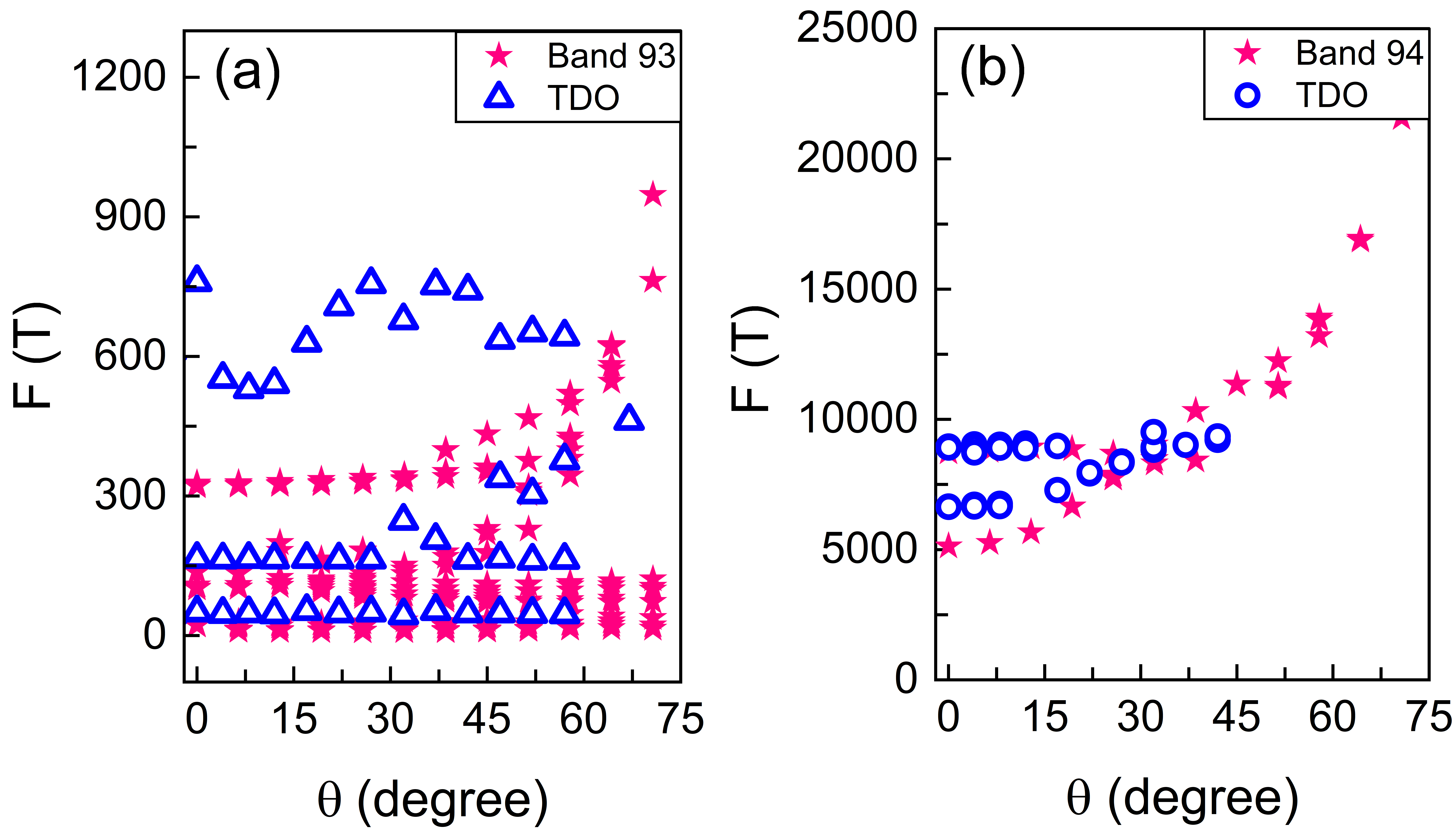}
\caption{Angular dependence of the experimental quantum-oscillation frequencies obtained from TDO measurements and the calculated frequencies associated with the Fermi-surface pockets of (a) Band 93 and (b) Band 94. The Fermi-surface pockets of Band 93 yield frequencies below $\sim300$ T, whereas those of Band 94 give frequencies of $\sim$5000 and 9000 T. The experimentally observed high-frequency branches are consistent with the calculated frequencies from Band 94, while the low-frequency branches near 54 and 150 T are consistent with those calculated from Band 93.}\label{Fig8}
\end{figure}

To compare the experimentally observed quantum-oscillation frequencies with the calculated Fermi surface, we calculated the extremal cross-sectional areas of the Fermi-surface pockets using the Onsager relation~\cite{shrestha2014shubnikov}. The calculations were performed using the SKEAF code~\cite{julian2012numerical}. Figure~\ref{Fig8} shows the angular dependence of the calculated frequencies associated with the Fermi-surface pockets of Band 93 and Band 94, together with the experimental frequencies obtained from TDO measurements. The Fermi-surface pockets of Band 93 yield relatively low frequencies, predominantly below $\sim300$ T, whereas Band 94 produces substantially higher frequencies, extending to $\sim$5000--9000 T. As shown in Fig.~\ref{Fig8}(b), the experimentally observed high-frequency branches are in good agreement with the frequencies calculated for Band 94, both in magnitude and angular dependence. The low-frequency branches near 54 and 150 T are also consistent with the calculated frequencies associated with Band 93 [Fig.~\ref{Fig8}(a)]. However, the experimentally observed frequencies near $\sim600$ T are not reproduced by the present DFT calculations. Although calculated frequencies near $\sim300$ T are present, they differ substantially from the experimental values. 

In general, discrepancies between experimental and calculated quantum-oscillation frequencies can arise from differences between the calculated and actual Fermi-surface topology, including possible shifts of the chemical potential relative to the DFT Fermi level~\cite{nguyen2022fermiology}. A rigid shift of the Fermi level is often considered when relatively small discrepancies between experimental and calculated frequencies are observed. In the present case, however, the experimental frequencies near $\sim600$ T are approximately twice the calculated frequencies near $\sim300$ T, making a simple rigid shift of the Fermi level an unlikely explanation for the discrepancy. It is also possible that some quantum-oscillation frequencies are not resolved in the present TDO measurements because of the limited signal quality and experimental sensitivity. Additional quantum-oscillation measurements using magnetization/torque magnetometry or Shubnikov--de Haas oscillations in magnetotransport would therefore be useful for further establishing the complete Fermi-surface topology and clarifying the origin of the $\sim600$ T frequencies.

\section{Summary and Conclusions}
We have investigated the magnetic and electronic properties of the kagome magnet HoV$_6$Sn$_6$ through magnetic susceptibility, field-dependent magnetization, TDO quantum-oscillation measurements, and first-principles calculations. The magnetic susceptibility exhibits Curie-Weiss behavior at high temperatures, with a Weiss temperature of $\theta_{\rm CW}=6.63(1)$ K and an effective magnetic moment of $\mu_{\rm eff}=11.11(1),\mu_B$/f.u., close to the expected free-ion value for Ho$^{3+}$. A magnetic transition occurs at $T_c\sim2.4$ K. The field-dependent magnetization at 2 K reveals pronounced magnetic anisotropy, with the magnetization approaching a saturation moment of $\sim9.5,\mu_B$/Ho for $H\parallel c$, whereas the moment reaches $\sim7.2,\mu_B$/Ho for $H\parallel ab$ without clear saturation up to 7 T. These results indicate a strong preference for the Ho moments to align along the $c$ axis. Our DFT+$U$ calculations further show that the ferromagnetic configuration with the Ho moments oriented along the $c$ axis is lower in energy than the nonmagnetic configuration, consistent with the ferromagnetic ground state and moment orientation established by previous neutron-scattering measurements~\cite{zhou2024ground}.

TDO measurements reveal multiple quantum-oscillation frequencies extending to $\sim9$ kT, demonstrating a complex multiband Fermi surface. The angular dependence of the oscillation frequencies provides information on the three-dimensional Fermi-surface topology, while the temperature dependence allows the cyclotron effective masses to be determined. First-principles calculations reveal multiple bands crossing the Fermi level, with the low-energy electronic states dominated by V-$d$ orbitals. The calculated Fermi surface gives quantum-oscillation frequencies that are broadly consistent with several of the experimentally observed branches. In particular, the low-frequency branches near 54 and 150 T and the high-frequency branches are consistent with the calculated frequencies associated with Band 93 and Band 94, respectively. However, experimental branches near $\sim600$ T are not reproduced by the present calculations. The magnitude of this discrepancy suggests that a simple rigid shift of the Fermi level is unlikely to provide a complete explanation. Further quantum-oscillation measurements, including torque magnetometry to probe de Haas--van Alphen oscillations and magnetoresistance measurements to detect Shubnikov--de Haas oscillations, would help resolve additional frequencies and provide a more complete determination of the Fermi-surface topology of HoV$_6$Sn$_6$. Taken together, the magnetic, quantum-oscillation, and first-principles results provide a comprehensive picture of the magnetic and electronic properties of HoV$_6$Sn$_6$ and establish a foundation for understanding the interplay between magnetism and electronic structure in kagome magnets.\\

\section*{acknowledgements}
This work was supported in part by the U.S. Department of Energy, Office of Science, Office of Workforce Development for Teachers and Scientists (WDTS) under the Visiting Faculty Program (VFP) at Los Alamos National Laboratory, administered by the Oak Ridge Institute for Science and Education. The work at West Texas A$\&$M University (WTAMU) is supported by the Killgore Undergraduate and Graduate Student Research Grants, the Welch Foundation (Grant No. AE-0025) and the National Science Foundation (Award No. 2336011). The computations were performed on the WTAMU HPC cluster, which was funded by the National Science Foundation (NSF CC* GROWTH 2018841). G.P. gratefully acknowledges the support of Dr. James ``Earl" Perry CMCS Fund and the University of West Georgia. S.D.W. gratefully acknowledges support via the UC Santa Barbara NSF Quantum Foundry funded via the Q-AMASE-i program under award DMR-1906325. A portion of this work was performed at the National High Magnetic Field Laboratory, which is supported by National Science Foundation Cooperative Agreement No. DMR-2128556 and the State of Florida, and the U.S. Department of Energy. 

\bibliography{Bibliography}

\newpage

\end{document}